\documentclass[cits]{PoS}

\newcommand{\SU}{\mathrm{SU}}

\title{Geometric effects in lattice QCD thermodynamics }

\ShortTitle{Geometric effects in lattice QCD thermodynamics }

\author{\speaker{Marco Panero}\\
        Institute for Theoretical Physics, University of Regensburg, 93040 Regensburg, Germany\\
        E-mail: \email{marco.panero@physik.uni-regensburg.de}}

\abstract{I present a study of the equation of state in quenched QCD, discussing some systematic effects related to the lattice geometry. In particular, I comment on the modification of the Stefan-Boltzmann law for a gas of free gluons in a finite system, and study the impact it might have on numerical results at high temperatures, for the typical parameters of current lattice simulations. Finally, I apply the results of this study to the analysis of data obtained from simulations of $\SU(N)$ gauge theories with $N>3$ colors, in a temperature range up to $3T_c$, where infrared effects appear to be under control. Preliminary results for various thermodynamic observables for $\SU(4)$, $\SU(5)$ and $\SU(6)$ gauge theories are found to be close to each other and to those for $\SU(3)$, in agreement with other similar studies. This may be relevant for the theoretical description of the QCD plasma.}

\FullConference{The XXVI International Symposium on Lattice Field Theory \\
		 July 14 - 19, 2008\\
		 Williamsburg, Virginia, USA}

\begin{document}

\section{Introduction and motivation}
\label{sec:intro}

Experiments at temperatures $T$ up to a few times the deconfinement temperature $T_c$ reveal a strongly interacting system~\cite{Adcox:2004mh}, which behaves as a nearly perfect fluid~\cite{Kolb:2003dz}. In this regime, lattice calculations are essentially the only tool to determine the equation of state (EoS) from the first principles of QCD. 

In particular, for the gluon sector of the QCD plasma, the lattice results show strong deviations from the Stefan-Boltzmann (SB) limit, with a large deficit in the entropy and pressure~\cite{Boyd:1996bx}. This led to conjecture~\cite{Gubser:1998nz} that the gluon plasma may admit an effective description based on the AdS/CFT correspondence~\cite{Maldacena:1997re}. Together with recent efforts to apply similar gauge/gravity techniques to build a holographic dual of QCD~\cite{Erlich:2005qh}, this has  triggered interest in lattice studies of $\SU(N)$ Yang-Mills thermodynamics with $N>3$ colors~\cite{Bringoltz:2005rr, Datta:2009tj, Panero:2009tv}.

One aspect of the $\SU(N>3)$ simulations is that, in general, they can be carried out using smaller lattices than those needed for $\SU(3)$---see, e.g. refs.~\cite{Lucini:2003zr}. This should not pose a problem of infrared (IR) effects, because for temperatures of the order of a few times $T_c$ finite-volume effects are expected to be exponentially suppressed, due to the existence of screening masses~\cite{DeTar:1985kx}. 

However, it was recently suggested~\cite{Gliozzi:2007jh} that nontrivial IR corrections affecting a gas of free gluons may still be relevant for the strongly interacting plasma at temperatures $O(T_c)$. For a periodic box of volume $L^3$ and timelike size $1/T$, these corrections depend logarithmically on the aspect ratio of a timelike cross-section $x:=LT$, and vanish for $x \rightarrow \infty$. In this contribution I study numerically the impact of these effects at temperatures up to about $3T_c$. Issues related to lattice IR effects were also studied in refs.~\cite{Elze:1988zs}.

\section{Finite-volume corrections to the partition function of the free gluon gas}
\label{sec:finite_volume_free_gluon_gas}

In the continuum, the partition function $\mathcal{Z}$ for a gas of free $\SU(N)$ gluons in a finite box of spacelike sizes $L \times L \times L$ at temperature $T$ was calculated exactly in ref.~\cite{Gliozzi:2007jh}:
\begin{equation}
\label{eq:partition_function}
\frac{\ln \mathcal{Z} }{N^2-1}=\frac{\pi^2}{45}(LT)^3-\ln\sqrt{LT}+
O(e^{-2\pi\,LT})\:.
\end{equation}
The leading finite-volume correction to $\ln \mathcal{Z}$ is a logarithmic function of $LT$. Neglecting terms $O(e^{-2\pi\,LT})$, the pressure $p$, energy density $\varepsilon$, and free energy density $f$ in a finite volume read: 
\begin{equation}
\label{eq:observables}
p = \frac{\varepsilon}{3} = \frac{\pi^2}{45} T^4 (N^2-1) \left[ 1 - \frac{15}{2 \pi^2(LT)^3} \right] \:,\;\; f = - \frac{\pi^2}{45} T^4 (N^2-1) \left[ 1 - \frac{45}{2 \pi^2(LT)^3} \ln (LT) \right] \:.
\end{equation}
Note that $p$ is no longer equal to $-f$, which is in contrast with the usual assumption underlying the determination of the EoS on the lattice with the integral method~\cite{Engels:1990vr}. The accuracy of these corrections for the $T \rightarrow \infty$ limit on the lattice is manifest when comparing them with, e.g., the finite-volume corrections to $\varepsilon$ obtained by numerical integration in ref.~\cite{Beinlich:1995ik}---see fig.~1 in ref.~\cite{Gliozzi:2007jh}.

One may wonder whether these effects can also play a r\^ole for a gas of strongly interacting gluons at finite temperature---in particular, at relatively low temperatures $O(T_c)$. As the logarithmic corrections appearing on the r.h.s. of eq.~(\ref{eq:partition_function}) stem from the regularization  of the divergent contribution to the functional integral coming from constant configurations, and periodic b.c. allow the existence of constant configurations also in the interacting system, it may be that a nontrivial IR correction still affects the gluon gas at relatively low temperatures. In particular, in ref.~\cite{Gliozzi:2007jh} it was pointed out that, for the typical parameters of present lattice simulations, this effect could account for a large fraction of the deviations from the SB limit observed at temperatures of the order of $T_c$.

\section{Numerical results}
\label{sec:results}

I run simulations of $\SU(N)$ gauge theories with the standard isotropic Wilson action, using an algorithm that combines heat-bath updates~\cite{Kennedy:1985nu} for $\SU(2)$ subgroups~\cite{Cabibbo:1982zn} and full-$\SU(N)$ overrelaxation updates~\cite{Kiskis:2003rd}; part of the simulations at $T=0$ were run using Chroma~\cite{Edwards:2004sx}. The lattice extent in the timelike direction in lattice units was $N_\tau = 5$ (and, in some cases, $6$); the spacelike volume in lattice units was $N_s^3$, with $N_s$ up to $22$, $18$, $16$ and $16$ for $\SU(3)$, $\SU(4)$, $\SU(5)$ and $\SU(6)$, respectively. The $T=0$ simulations were run on $N_s^4$ lattices. 

For $\SU(3)$, I set the scale using $r_0$~\cite{Necco:2001xg}, while for the other groups I used the string tension~\cite{Lucini:2004my}. I determined the EoS using the integral method~\cite{Engels:1990vr}, from measurements of the average plaquette at finely separated $\beta$ values: in particular, $n=180$ intervals of amplitude $\Delta \beta=0.005$ were used for $\SU(3)$. The numerical integration was done with the method described by eq.~(A.4) in ref.~\cite{Caselle:2007yc}, where it was used to measure the interface tension in the Ising model\footnote{See refs.~\cite{Caselle:2002ah} for further details.}; the errors of this method are $O(n^{-4})$. Cutoff effects on the asymptotic normalization of the EoS were corrected using the $R_I(N_\tau)$ factor~\cite{ Bringoltz:2005rr, Engels:1999tk}.

The IR corrections to the EoS can be calculated analytically only for the free-gluon gas; the result obtained in this limit can be considered as an upper bound for possible nontrivial IR effects in the strongly interacting theory at temperatures of the order of $T_c$. To check if in this regime the system is sensitive to logarithmic finite-volume corrections, I calculated the pressure from lattices characterized by different values of $LT$ in two different ways. First, assuming that the data are not affected by nontrivial IR effects. Second, assuming maximal sensitiveness, i.e., assuming the same corrections as for the free-gluon gas.\footnote{Since the quantity that is measured on the lattice is $-f$, rather than $p$, the latter method requires that the data be shifted (to compensate for $p+f \neq 0$) and rescaled (to account for the different high-temperature limit).} If the IR corrections affecting the gas of free gluons were still relevant at $T \simeq 2T_c$ or $3T_c$, then the latter method should give better consistency among results obtained from different lattices.

\begin{figure}[-t]
  \includegraphics[width=.25\textwidth,angle=270]{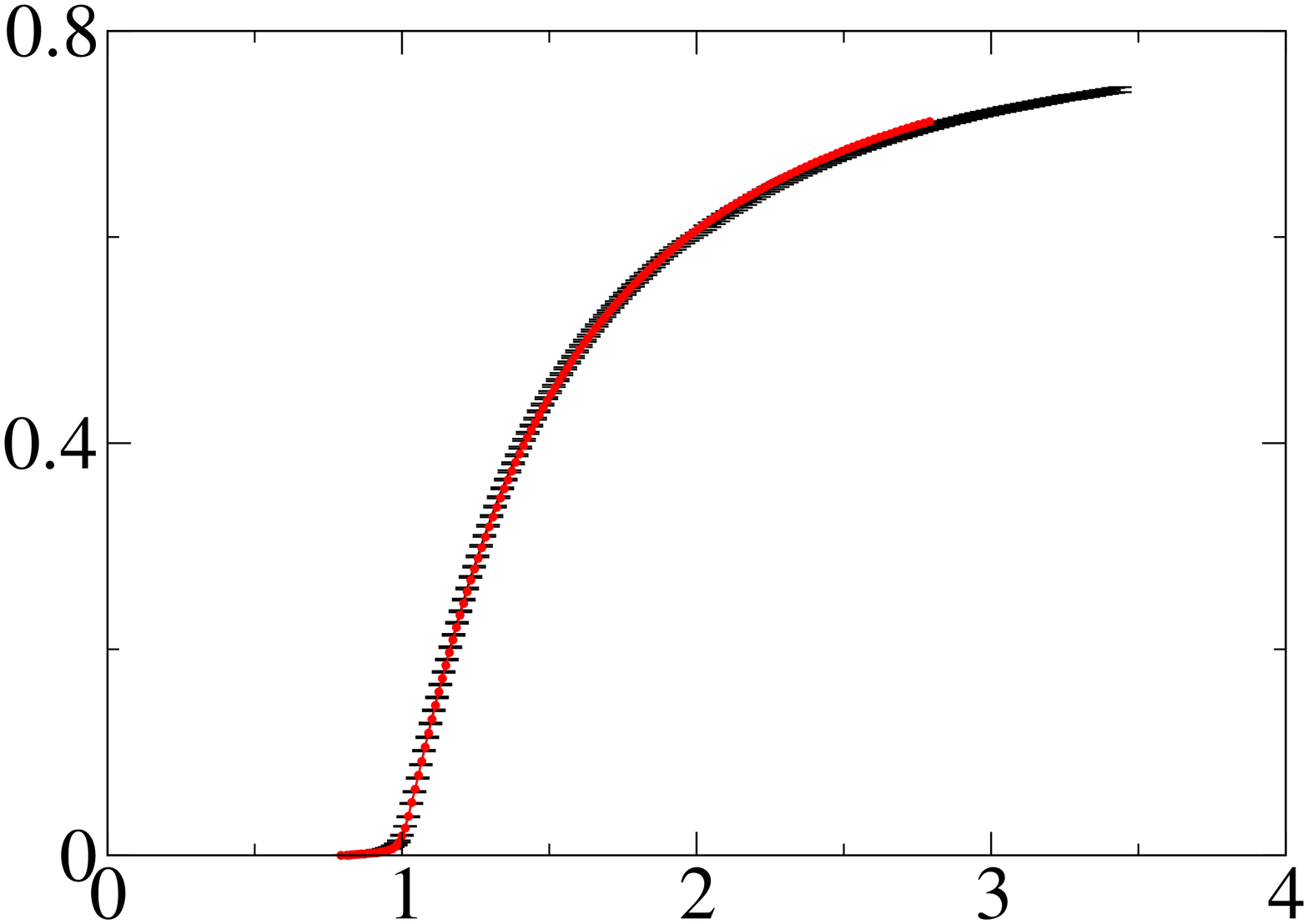}
  \includegraphics[width=.25\textwidth,angle=270]{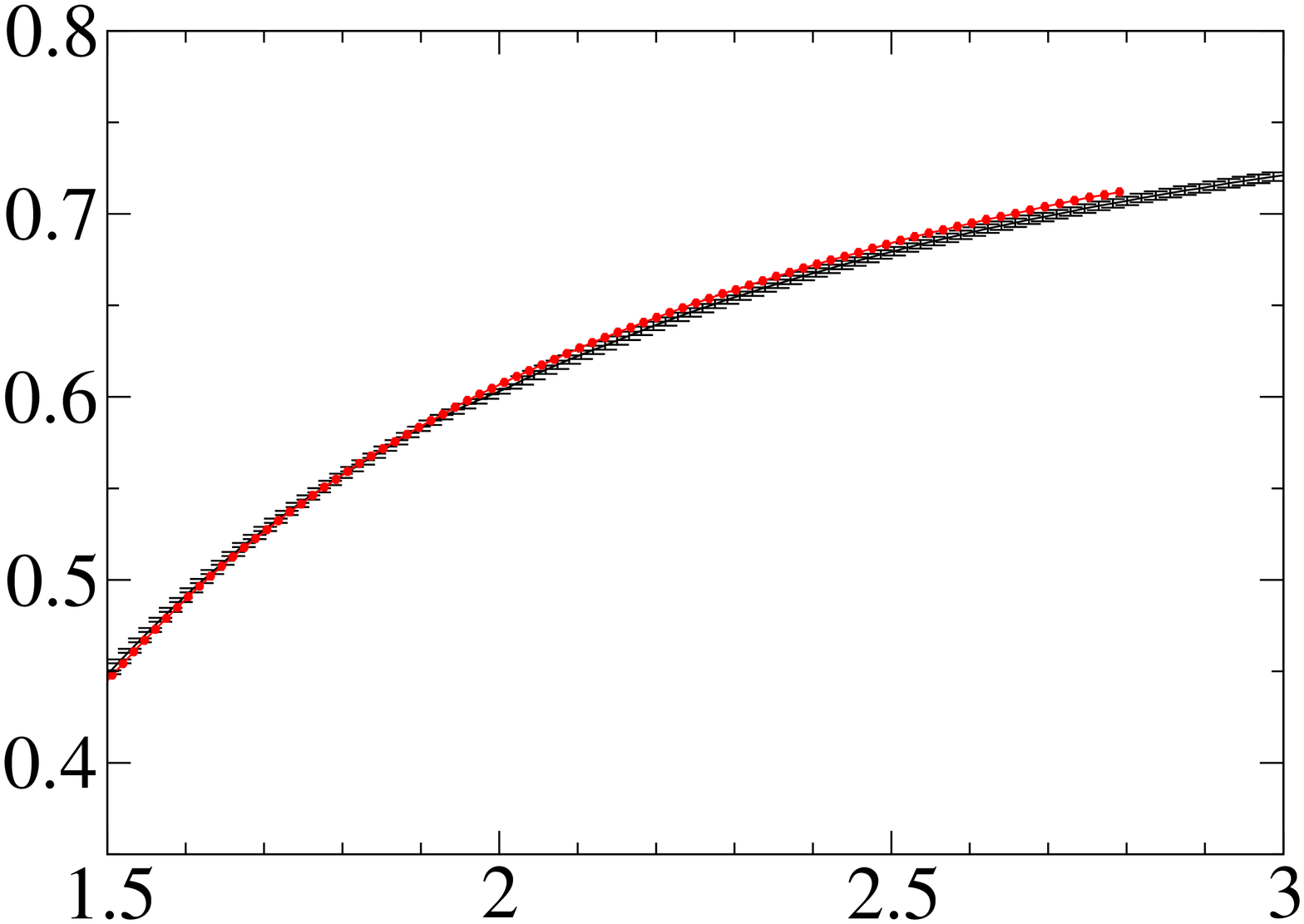}
  \includegraphics[width=.25\textwidth,angle=270]{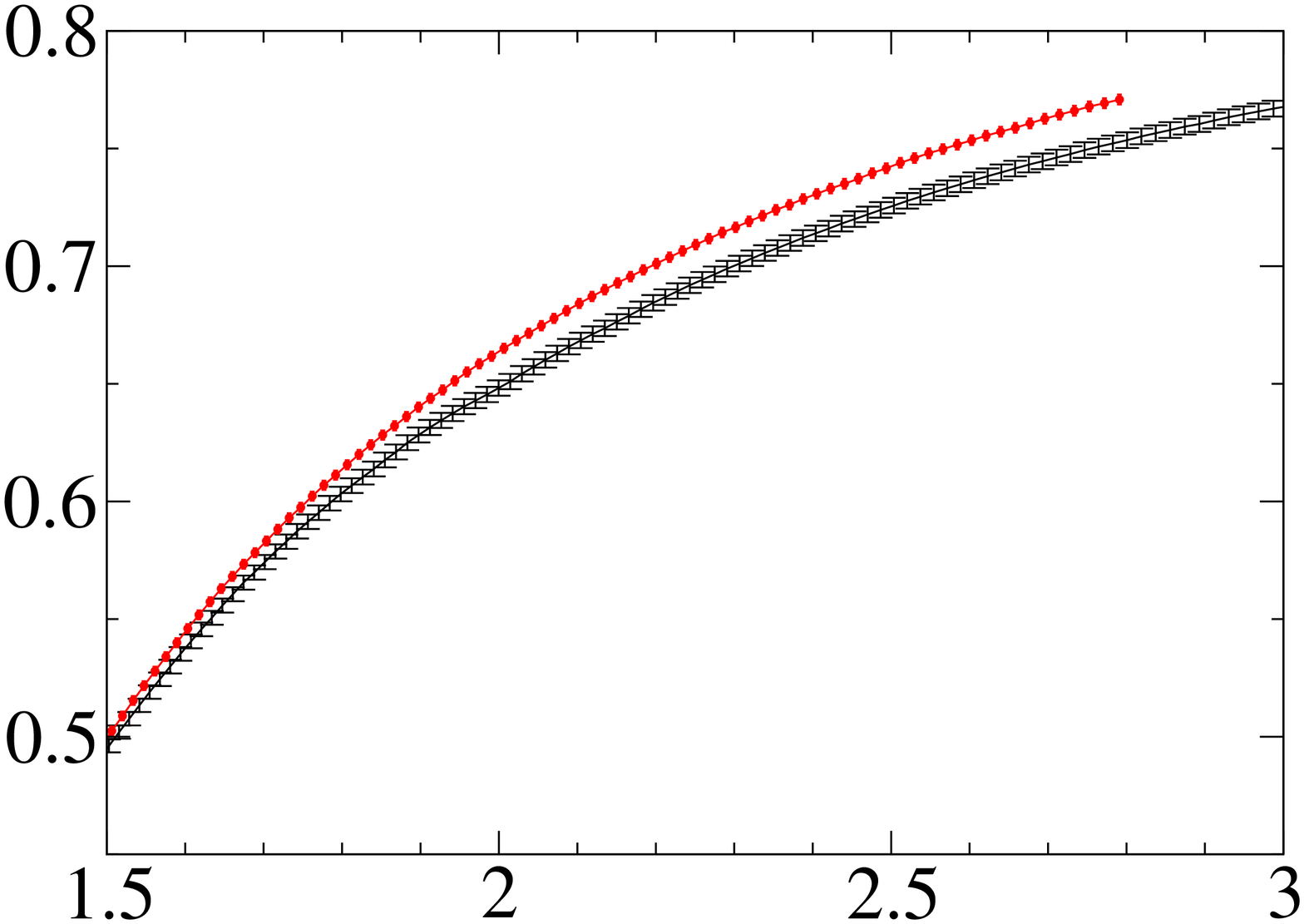}
  \caption{Left panel: The EoS for $\SU(3)$, evaluated  neglecting possible logarithmic corrections to the partition function, for lattices of size $20^3 \times 5$ (black crosses) and $18^3 \times 5$ (red circles). The plot shows the results for $p/T^4$ (normalized to the SB limit) \emph{vs.} $T/T_c$. Central panel: High-temperature zoom of the previous plot. Right panel: The EoS evaluated assuming maximum sensitiveness to IR effects (i.e. compensating for the logarithmic effects that affect a free-gluon gas), according to eq.~(\protect\ref{eq:observables}).} \label{fig:su3_pressure_threeplots}
\end{figure}

This, however, seems not to be the case: the left panel of fig.~\ref{fig:su3_pressure_threeplots} shows that the discrepancies between the results obtained using the first method are compatible with statistical fluctuations; this still holds up to $T \simeq 3T_c$ (central panel). On the contrary, the second method overcompensates the differences between the two data sets, driving the curves away from each other (right panel). Similar results also hold for $\SU(4)$.

Therefore, within the precision of our data, at temperatures up to $T \simeq 3 T_c$ we do not observe evidence for the logarithmic IR corrections that affect the free-gluon gas. Yet, as these corrections should eventually show up in the perturbative regime at high enough temperatures, it would be interesting to study at what $T$ values they can be seen on the lattice. Since the logarithmic corrections are nonnegligible only for $LT=O(1)$, this problem could be studied by running lattice simulations at large $\beta$ values and shrinking physical volume.\footnote{The $\SU(3)$ EoS at very high temperatures was studied in ref.~\cite{Endrodi:2007tq}, where the results shown in fig.~5 were obtained from simulations with $N_\tau=8$ and $N_s^3=24^3$ or $N_s^3=32^3$~\cite{Szabo_private_communication}. The maximal IR correction would then shift the asymptotic value down to about $97\%$ of the SB limit, which indeed looks compatible with their result at the highest temperature $T = 3 \times 10^7 T_c$ (although the data are also compatible with the conventional SB limit). Note however, that this comparison is only approximate, because in the numerical method of ref.~\cite{Endrodi:2007tq} the geometric parameter $x$ is not held fixed, and furthermore the pressure at high temperatures is obtained using the integral method, starting from low temperatures. This is based on the $p=-f$ equality, which---as we have seen---is reliable in the low-$T$ regime, but would also start being affected by IR corrections when $T$ is increased.} Although such a limit has its own reasons of theoretical interest~\cite{Bruckmann:2008xr}, in the present work I restricted my analysis to temperatures $O(T_c)$, and to sufficiently large physical volumes.

\begin{figure}[-t]
  \includegraphics[width=.25\textwidth,angle=270]{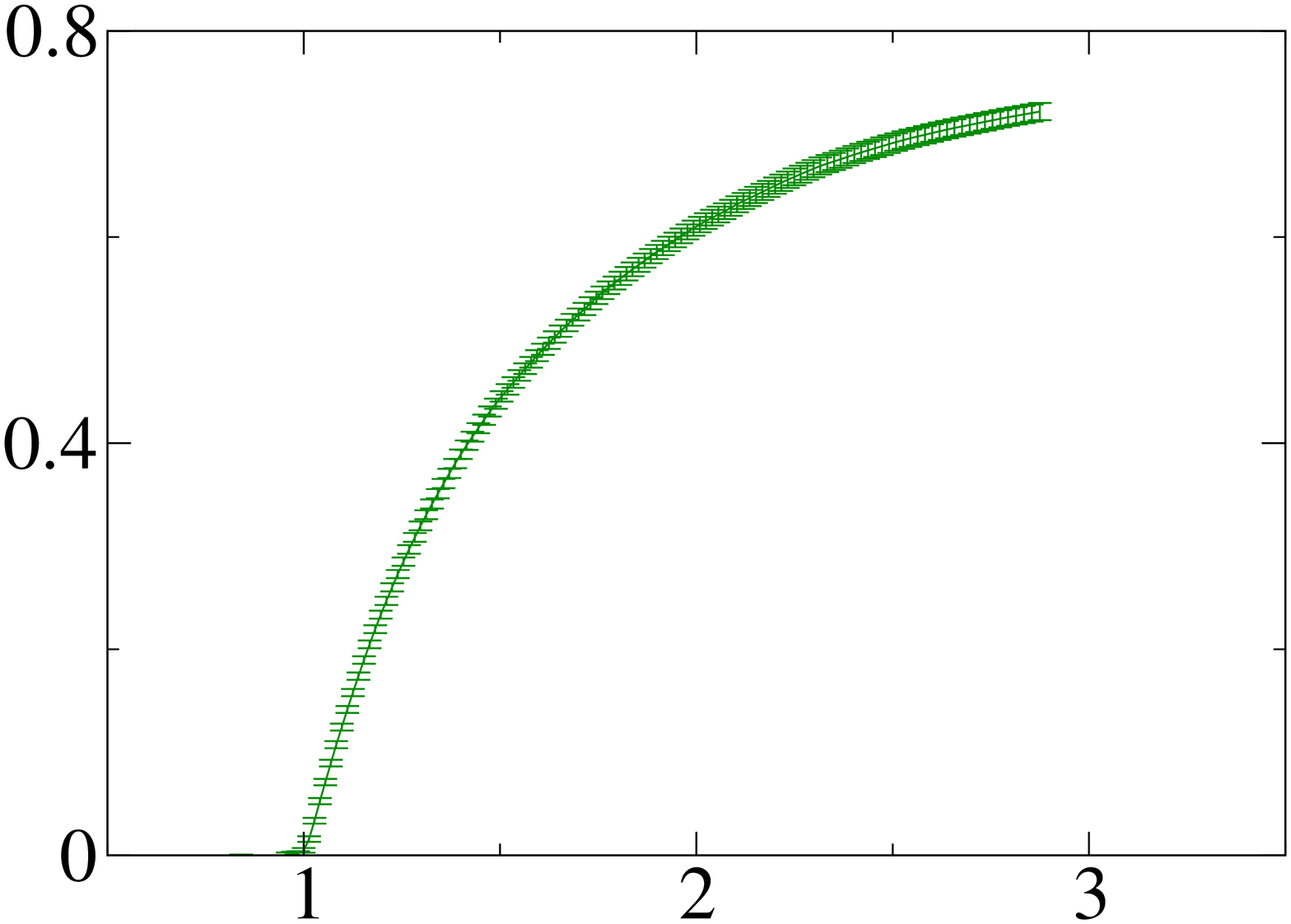}
  \includegraphics[width=.25\textwidth,angle=270]{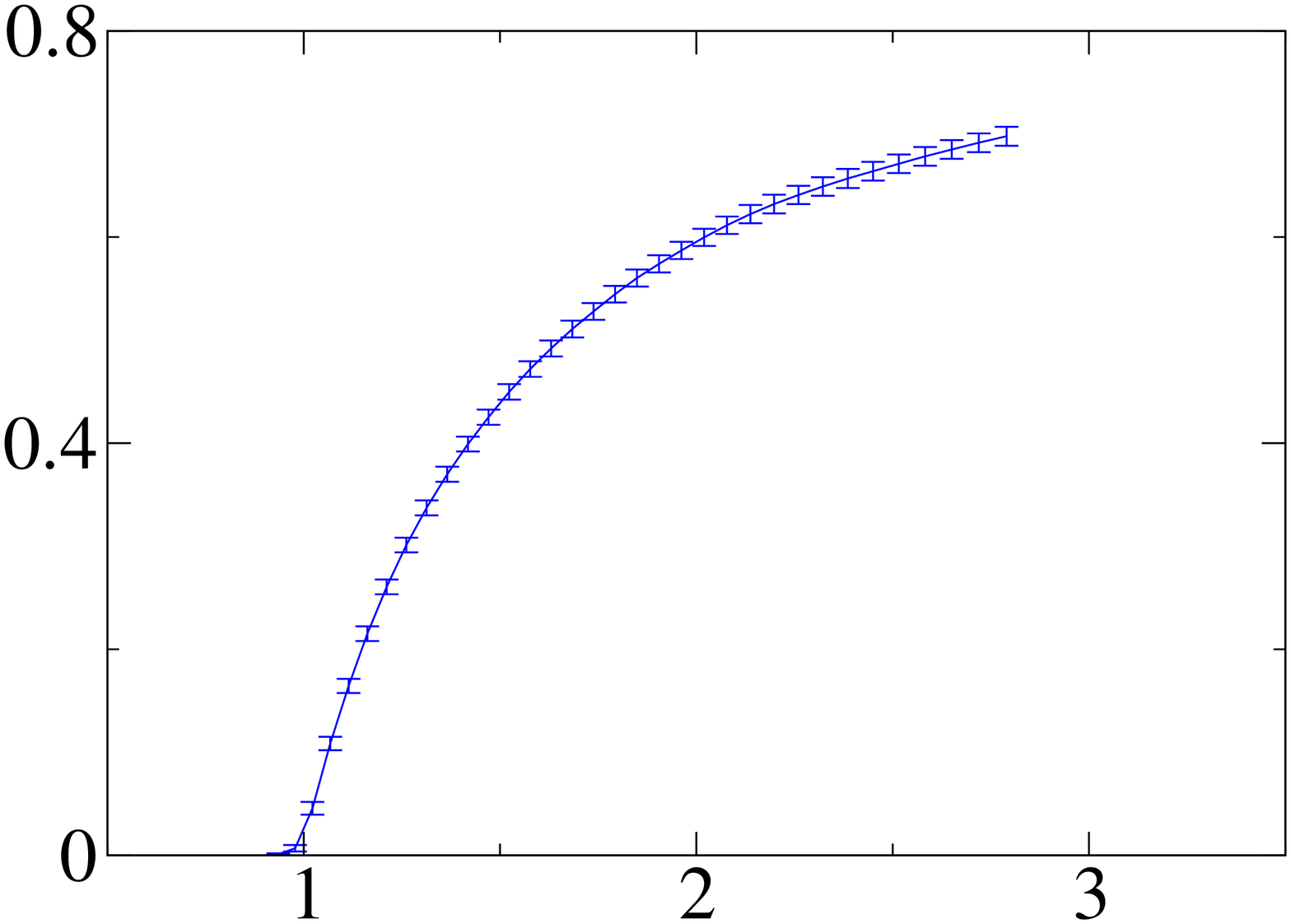}
  \includegraphics[width=.25\textwidth,angle=270]{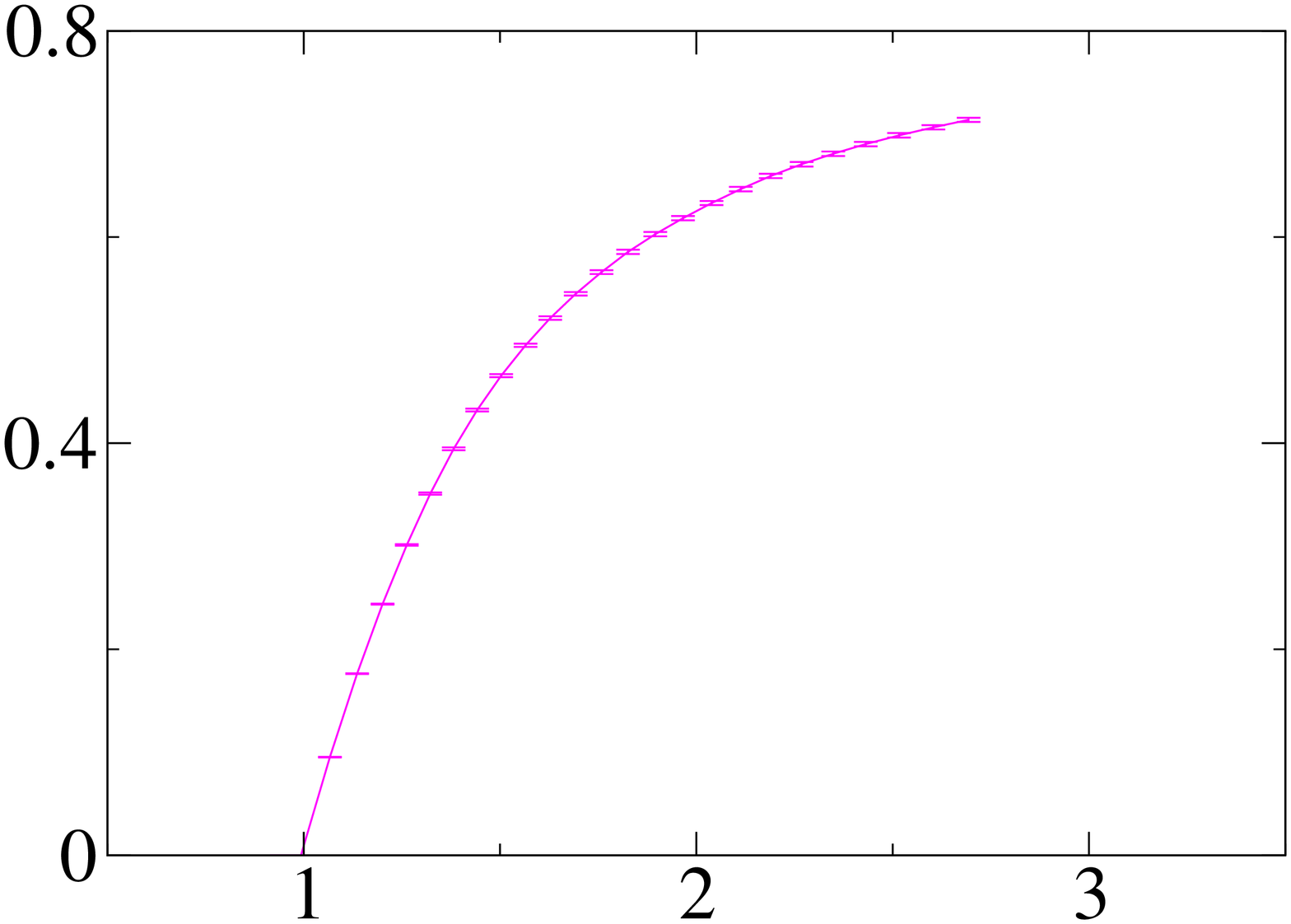}
  \caption{The rescaled pressure $p/T^4$ (normalized to the SB limit) from simulations on a $16^3 \times 5$ lattice, \emph{vs.} $T/T_c$, for $\SU(4)$ (left panel), $\SU(5)$ (central panel) and $\SU(6)$ (right panel).} \label{fig:pressure_various_groups}
\end{figure}

\begin{figure}[-t]
  \includegraphics[width=.19\textwidth,angle=270]{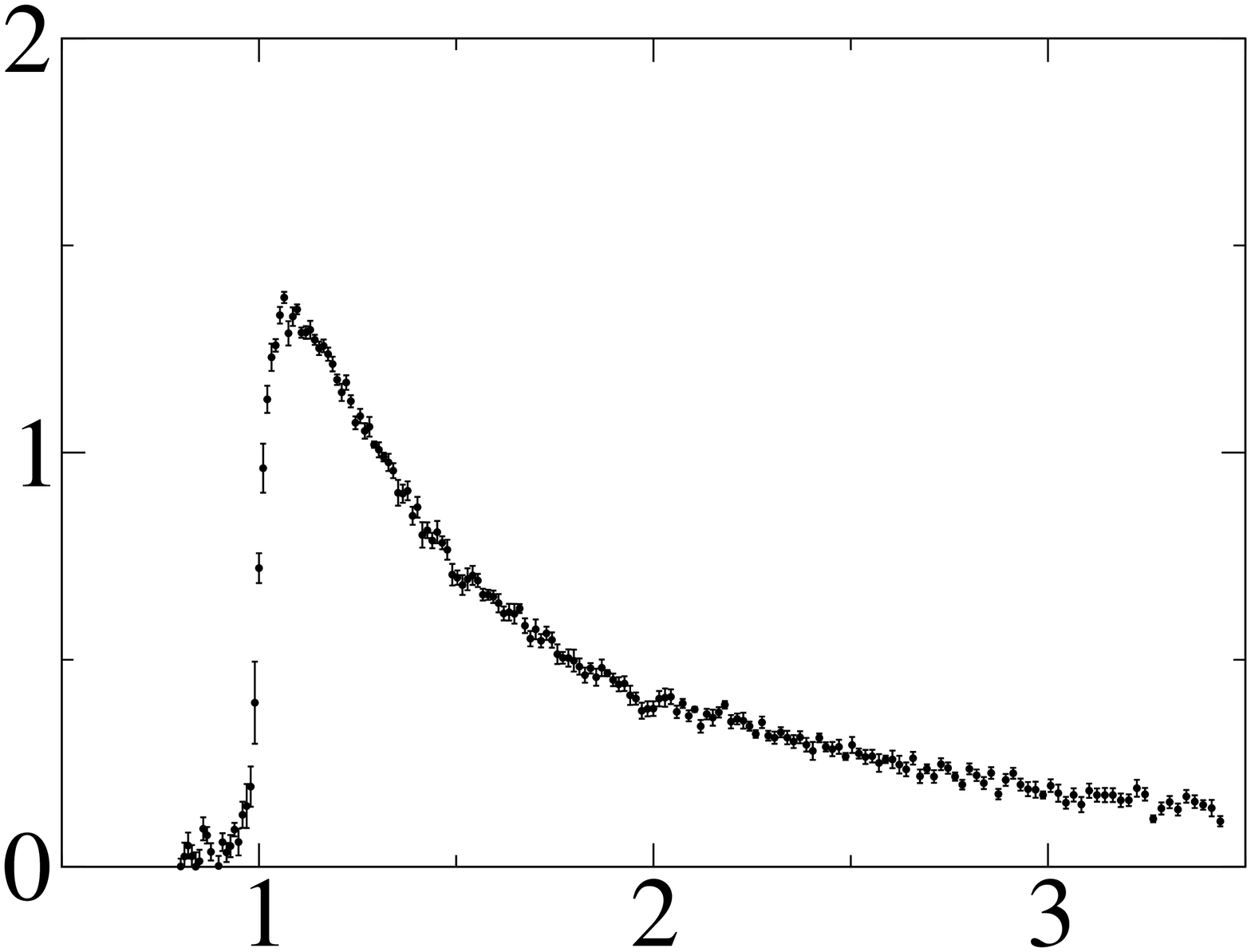}
  \includegraphics[width=.19\textwidth,angle=270]{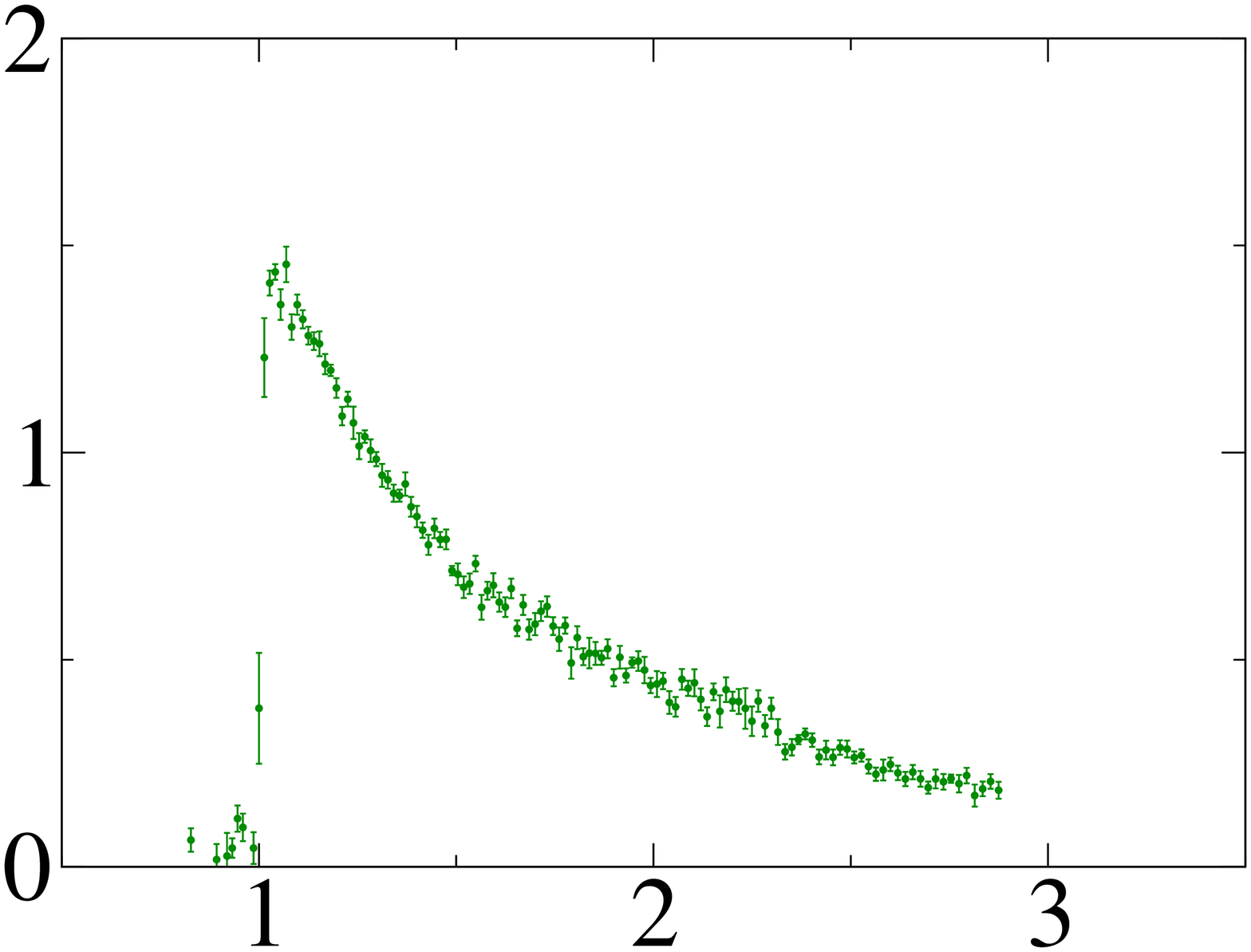}
  \includegraphics[width=.19\textwidth,angle=270]{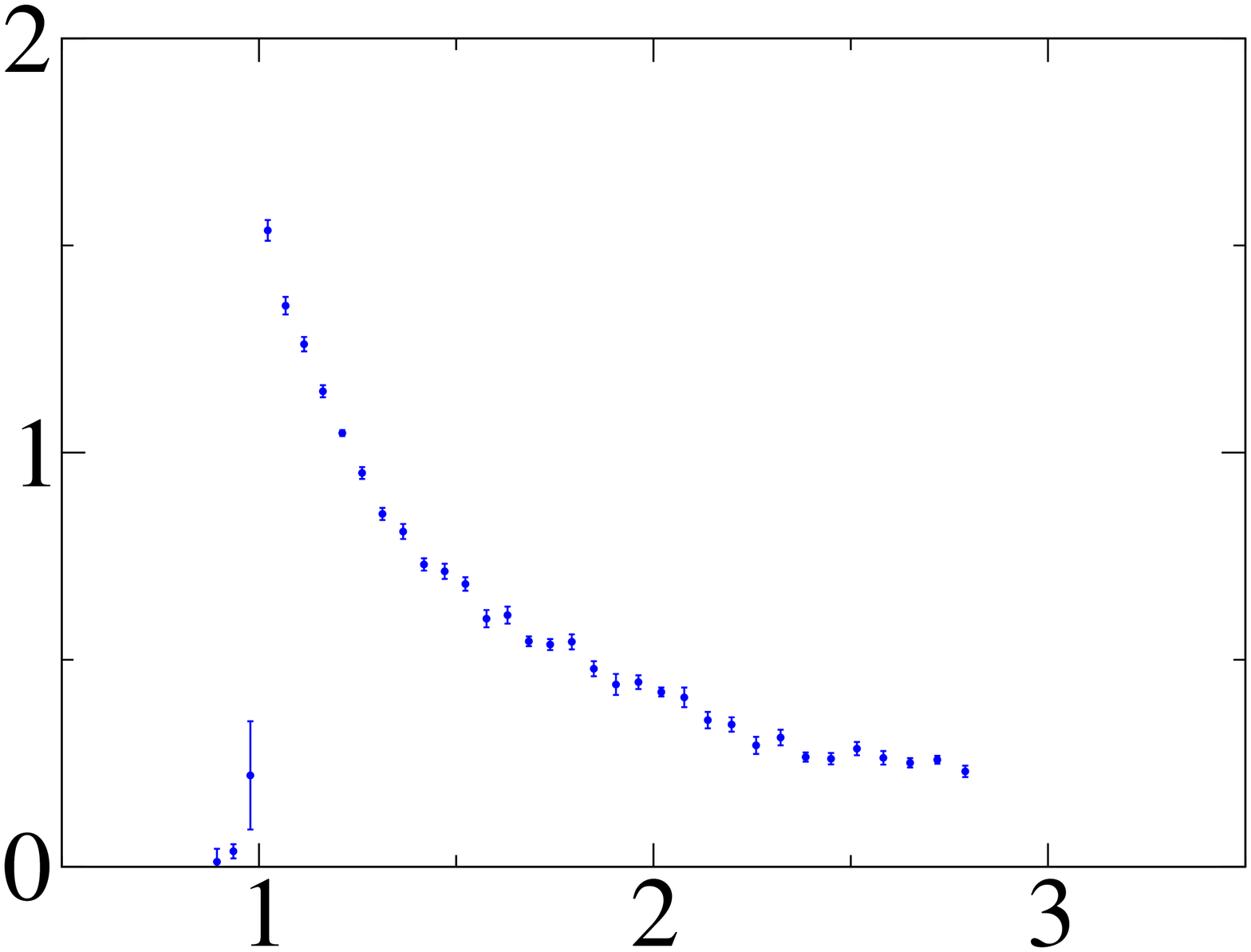}
  \includegraphics[width=.19\textwidth,angle=270]{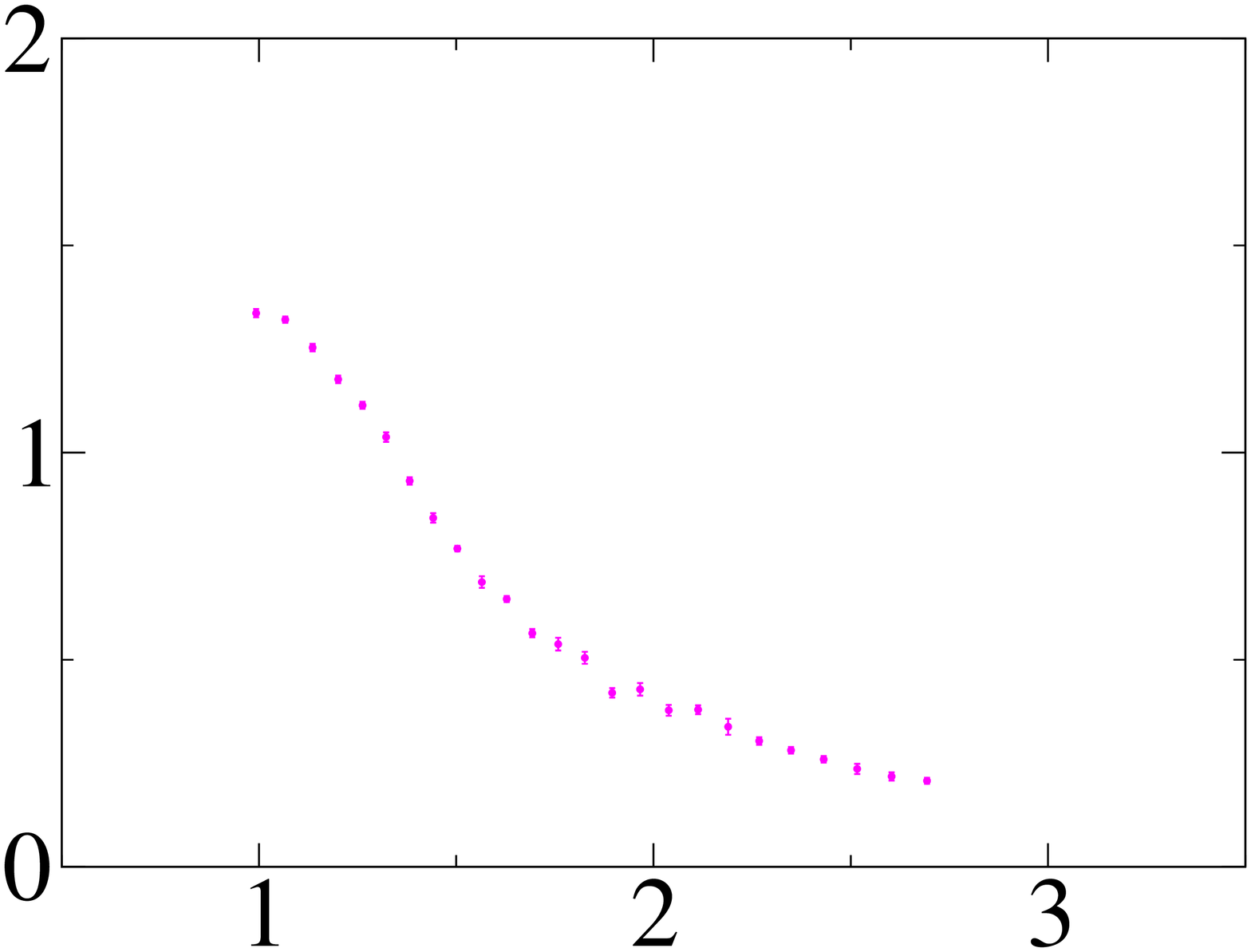}
  \caption{Left to right: $\Delta/T^4$, normalized to $\frac{\pi^2}{45}(N^2-1)$, \emph{vs.} $T/T_c$, for $\SU(3)$, $\SU(4)$, $\SU(5)$  and $\SU(6)$.} \label{fig:rescaled_trace_various_groups}
\end{figure}

\begin{figure}[-t]
  \includegraphics[width=.19\textwidth,angle=270]{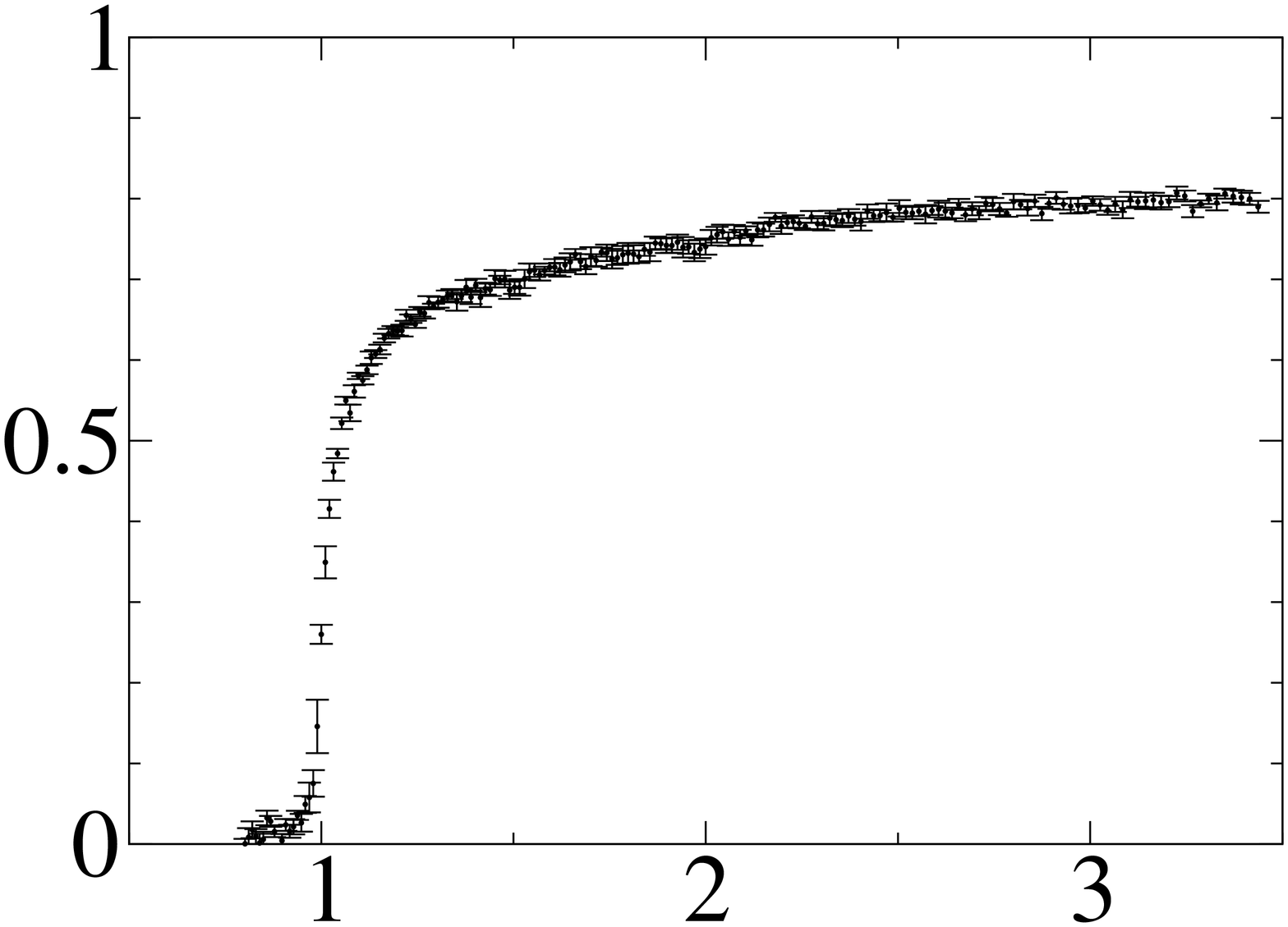}
  \includegraphics[width=.19\textwidth,angle=270]{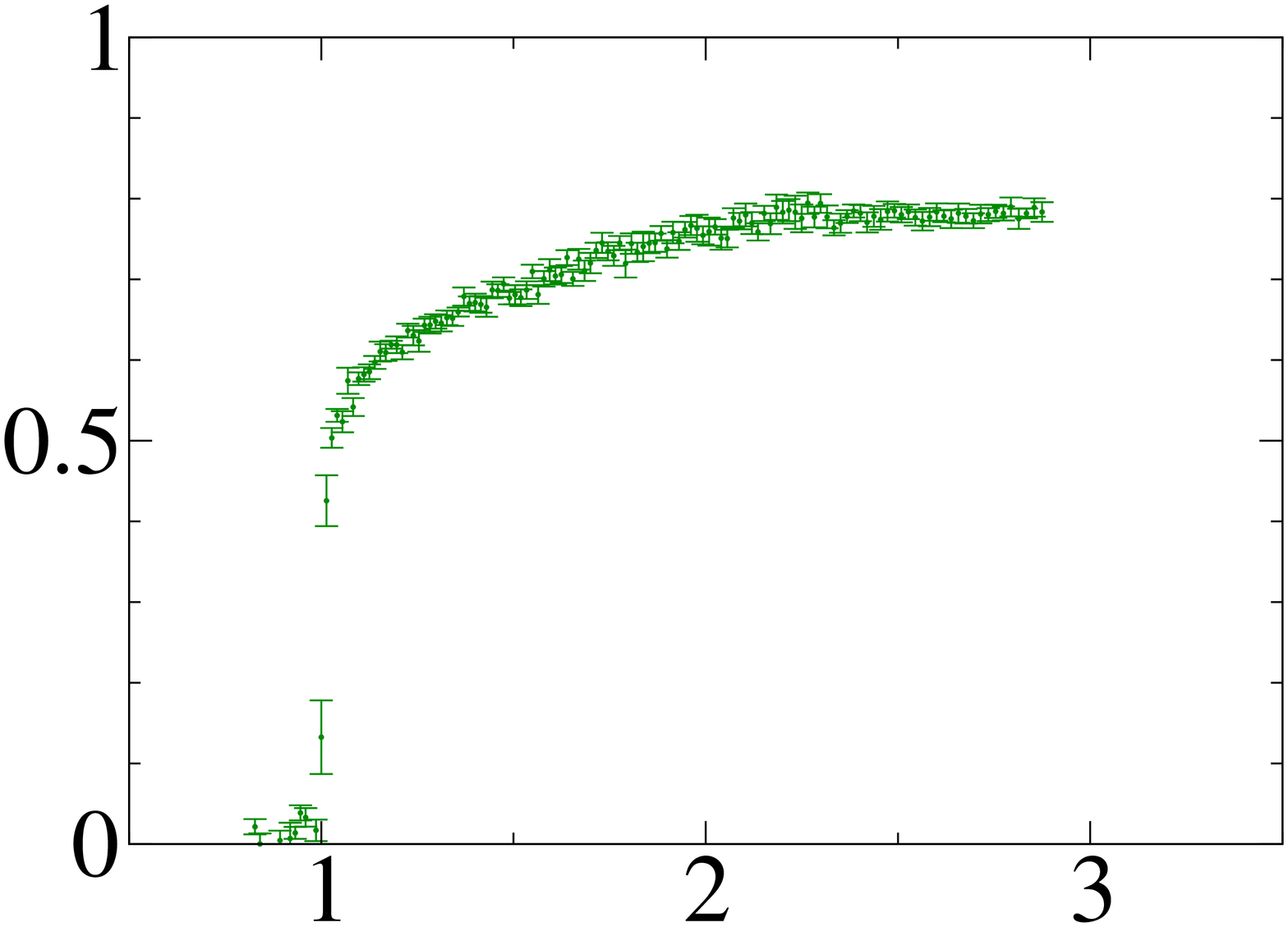}
  \includegraphics[width=.19\textwidth,angle=270]{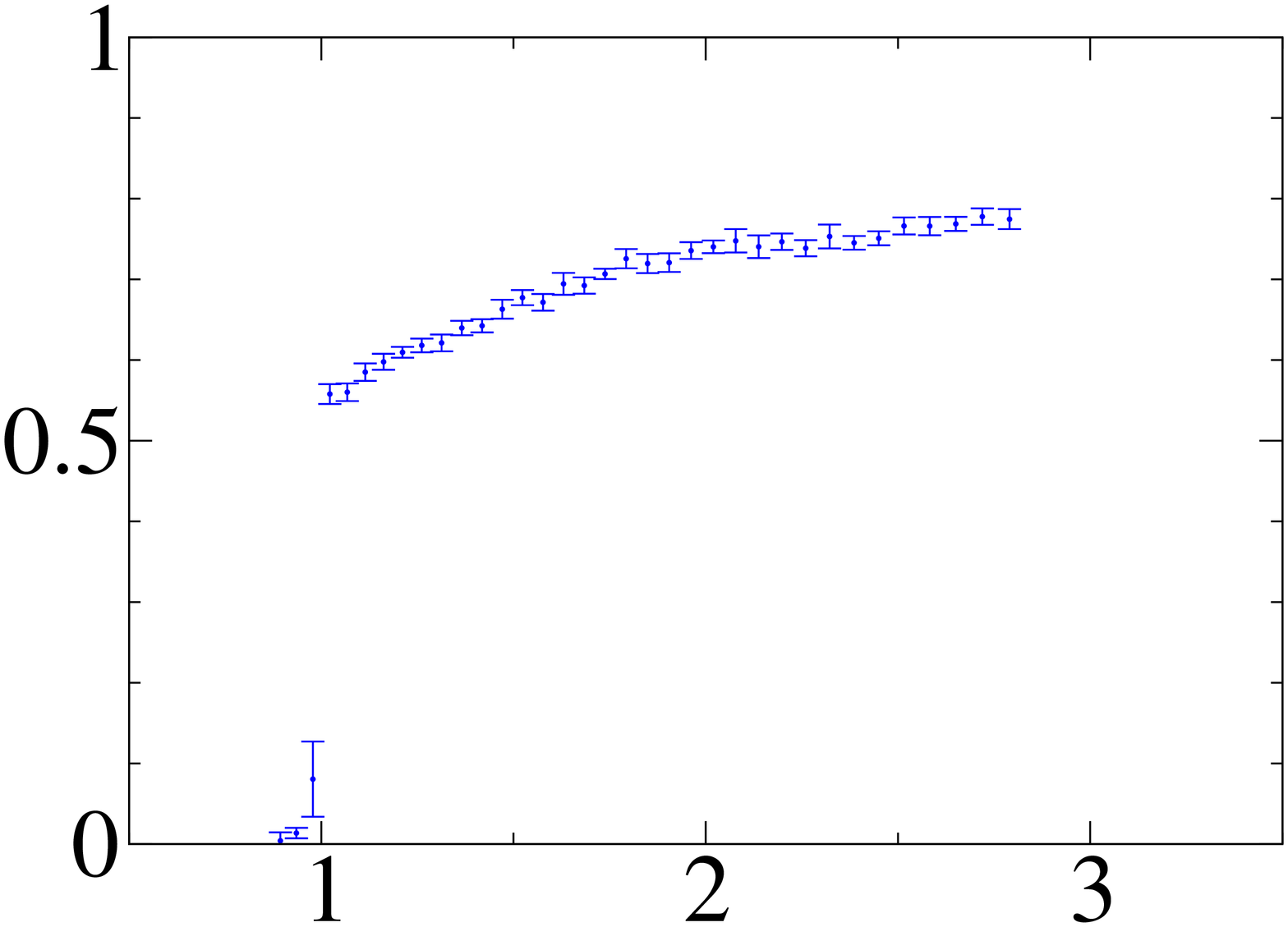}
  \includegraphics[width=.19\textwidth,angle=270]{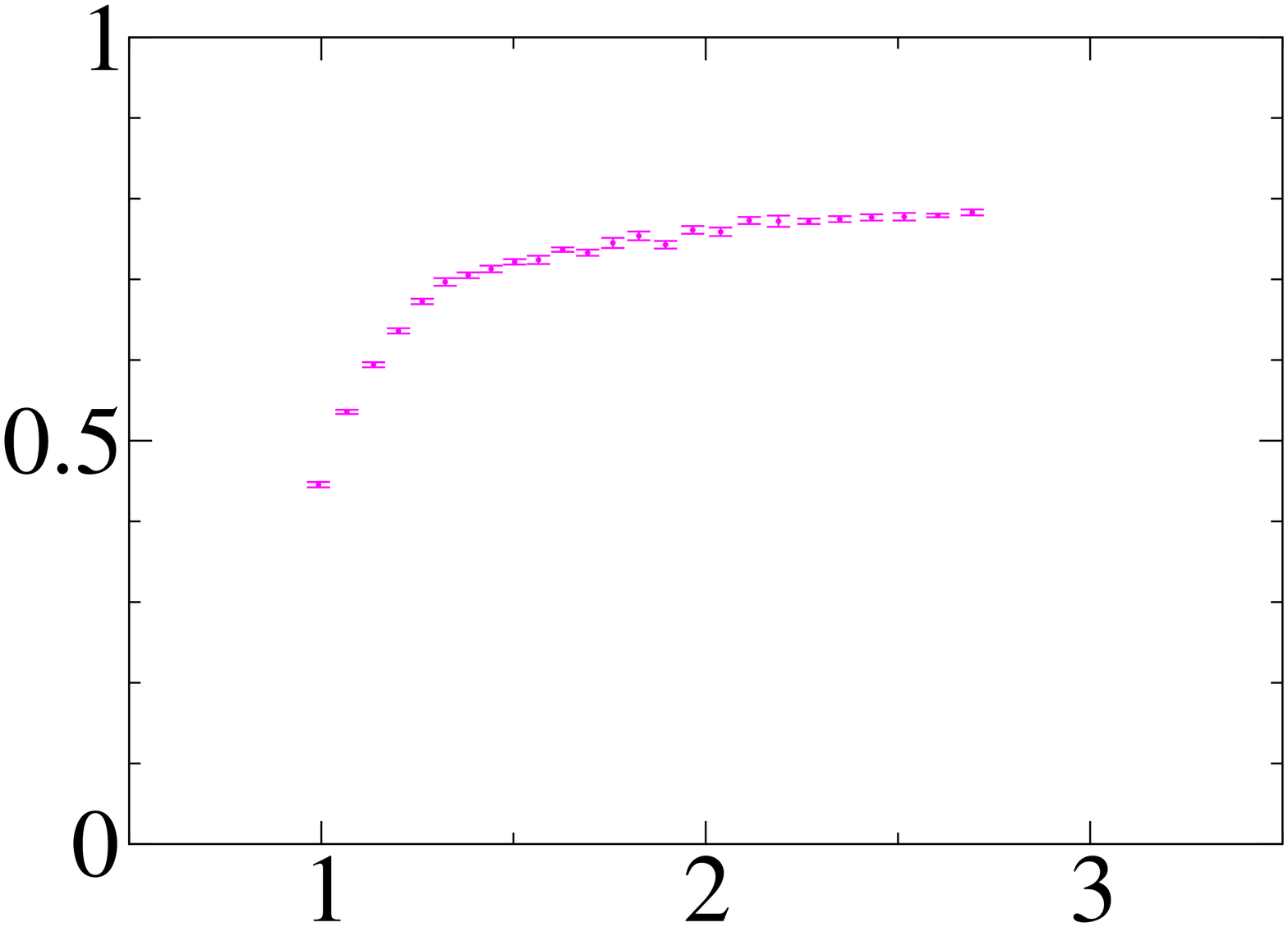}
  \caption{Left to right: The rescaled energy density $\varepsilon/T^4$ (normalized to its SB limit) \emph{vs.} $T/T_c$, for $\SU(3)$, $\SU(4)$, $\SU(5)$  and $\SU(6)$.} \label{fig:energy density_various_groups}
\end{figure}

\begin{figure}[-t]
  \includegraphics[width=.19\textwidth,angle=270]{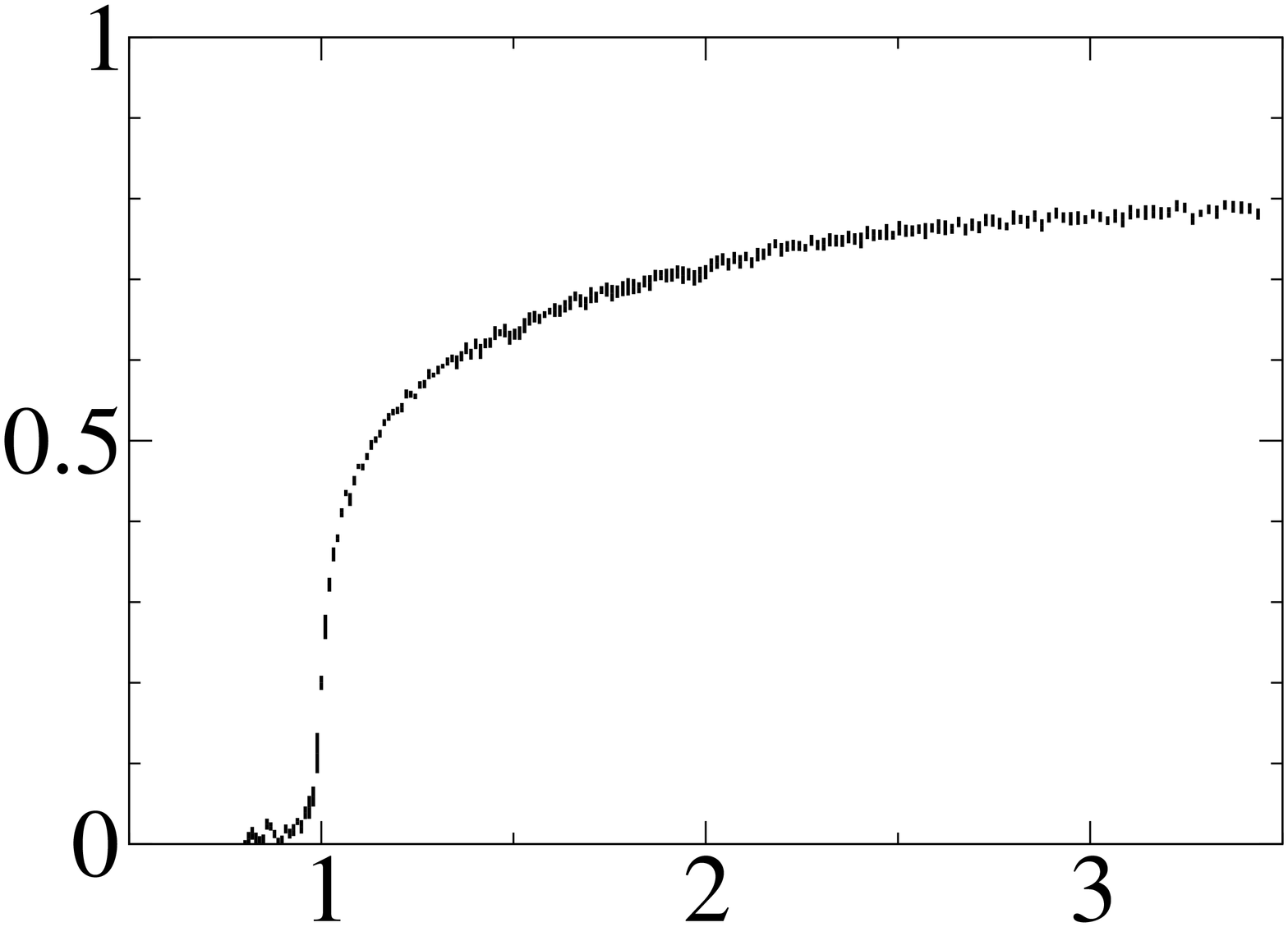}
  \includegraphics[width=.19\textwidth,angle=270]{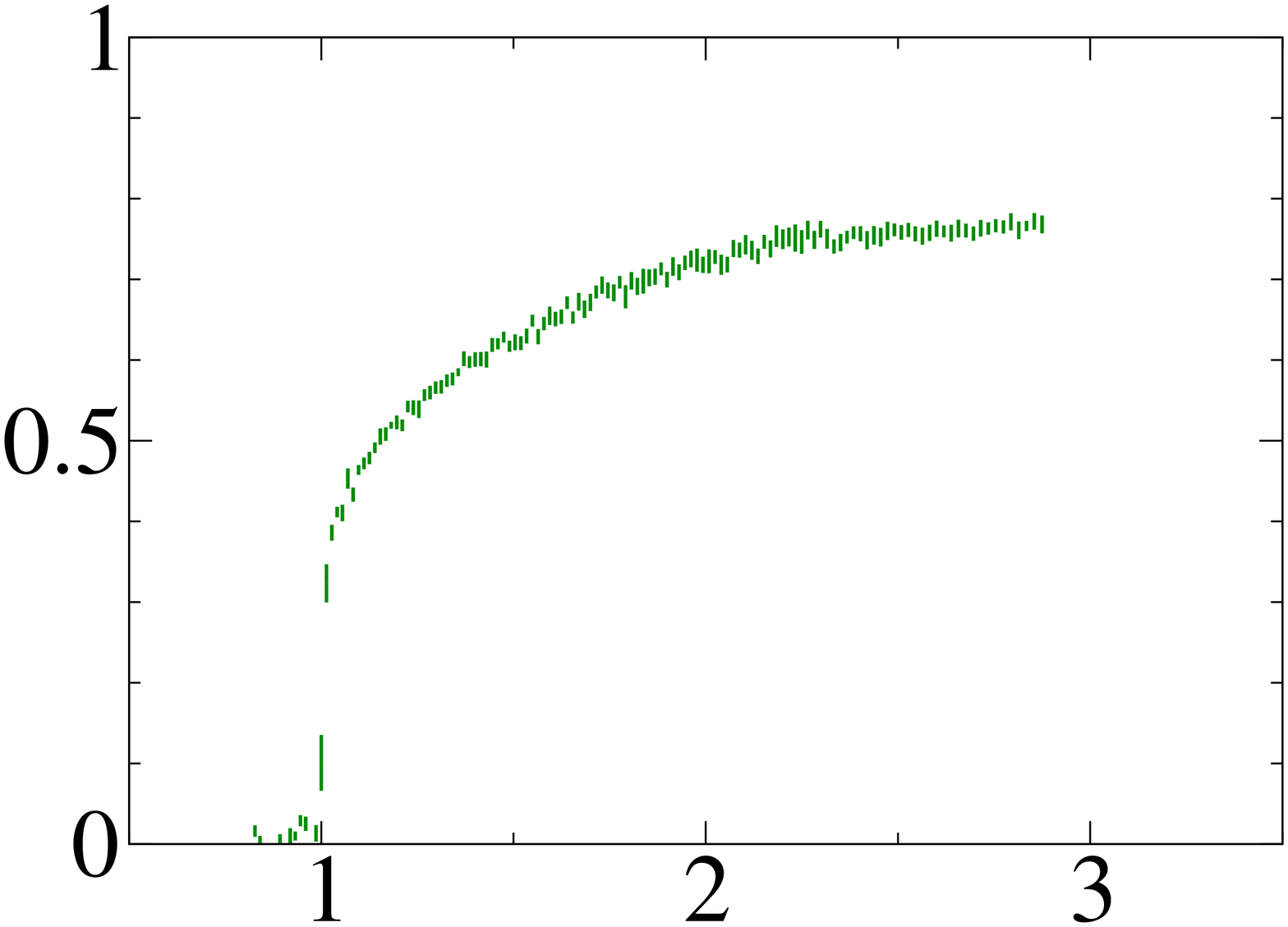}
  \includegraphics[width=.19\textwidth,angle=270]{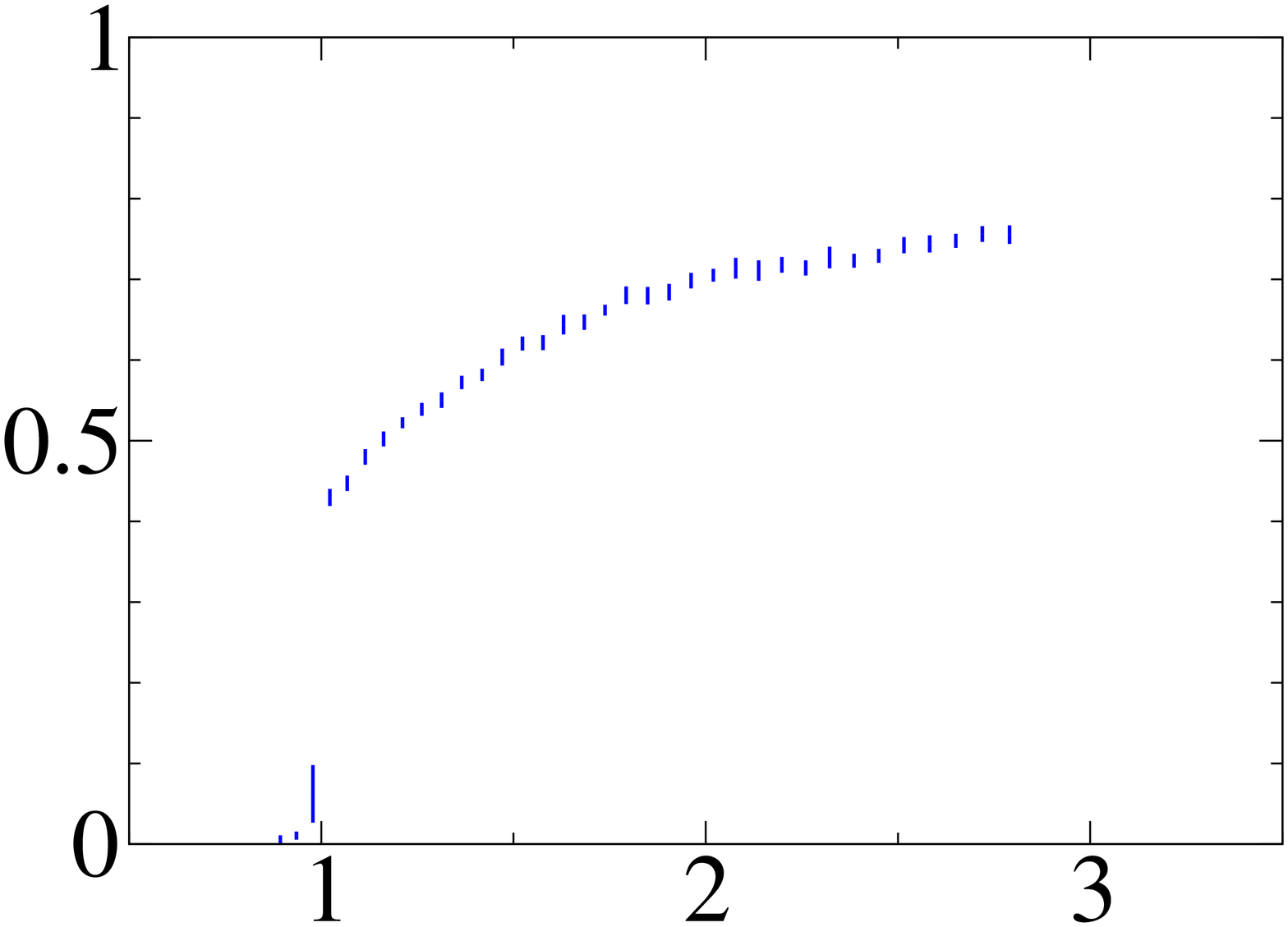}
  \includegraphics[width=.19\textwidth,angle=270]{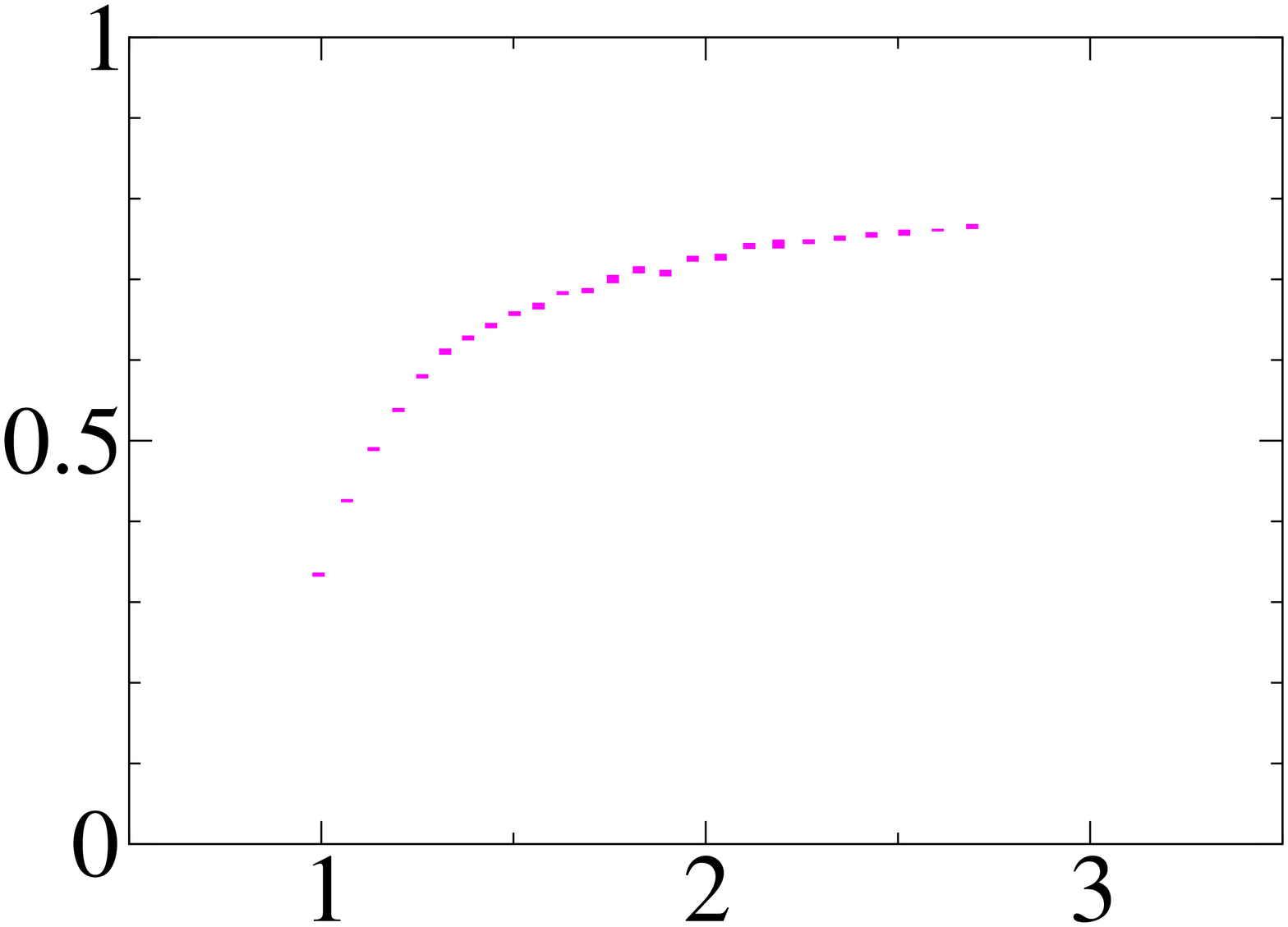}
  \caption{Same as in fig.~\protect\ref{fig:energy density_various_groups}, but for the rescaled entropy density $s/T^3$.} \label{fig:entropy_density_various_groups}
\end{figure}

Having found no evidence for logarithmic IR corrections in the $T<3T_c$ regime, I proceeded to the evaluation of thermodynamic observables in $\SU(N)$. Figs.~\ref{fig:pressure_various_groups}-\ref{fig:entropy_density_various_groups} show preliminary results for equilibrium thermodynamics observables (pressure, trace of the stress tensor $\Delta = \epsilon - 3p $, energy density and entropy density $s$), for $3 \le N \le 6$, normalized to their SB limits. 
In agreement with the conclusions of ref.~\cite{Bringoltz:2005rr},
the results from various gauge groups are essentially compatible with each other (with small quantitative corrections), and clearly different from the SB limit. Further results, including data for the $\SU(8)$ gauge group, are presented elsewhere~\cite{Panero:2009tv}.

\section{Summary and outlook}
\label{sec:conclusions}

The nontrivial IR corrections affecting a gas of free gluons in a finite box were derived in ref.~\cite{Gliozzi:2007jh}, where it was also suggested that they may still play a r\^ole at temperatures of the same order of magnitude as $T_c$. To the level of precision achieved, my numerical results in a (phenomenologically relevant) temperature range up to about $3T_c$ did not reveal any nontrivial finite-volume effects. However, the latter might probably be observed in numerical simulations at very high temperatures and small volumes~\cite{Endrodi:2007tq}. Next I also measured bulk thermodynamic quantities in $\SU(N)$ gauge theories with $3 \le N \le 6$ colors, at temperatures up to $3T_c$. In agreement with ref.~\cite{Bringoltz:2005rr}, the results for different $\SU(N)$ groups share similar features, and the strong deviations from the SB limit observed in $\SU(3)$ survive the large-$N$ limit. This leans support to the hypothesis that QCD may admit a nearly conformal effective description at temperatures of a few times $T_c$, where the theory is still strongly interacting~\cite{Gavai:2004se}.

\acknowledgments

I acknowledge support from the Alexander~von~Humboldt Foundation, and thank G.~S.~Bali, Ph.~de~Forcrand, F.~Gliozzi, S.~Gupta, B.~Lucini, K.~K.~Szab\'o and G.~D.~Torrieri for discussions. The University of Regensburg hosts the SFB/TR 55 ``Hadronenphysik mit Gitter-QCD''.

\end{document}